\documentclass{DISproc}

\begin{document}
\title{
\raggedright
Measurements of dibosons with the ATLAS detector and associated constraints on new physics}

\author{\raggedright {\slshape Hugh Skottowe}\\
skottowe@physics.harvard.edu\\
Harvard University,
18 Hammond Street,
Cambridge MA 02138, USA\\[1ex]
{\itshape on behalf of the ATLAS collaboration}}

\contribID{xy}

\doi  

\maketitle

\begin{abstract}
Diboson cross sections have been measured for all combinations of W,Z and isolated photons,
using the ATLAS detector at the LHC.
The cross sections are measured in kinematic regions defined by the decay kinematics,
in some cases including vetoes on additional jets.
The measurements are also extrapolated to the full phase space using
theoretical calculations of the acceptance, and are additionally used
to place constraints on triple-gauge boson couplings.
\end{abstract}

\section{Introduction}

At the end of the 2011 proton-proton run of the LHC, the delivered luminosity
was increased by a factor of $\sim$5 from the $\sim$1\,fb$^{-1}$ dataset
used for summer 2011 conference results.
New measurements using the full $\sim$5\,fb$^{-1}$ dataset are presented here.
The larger dataset implies a significant reduction of
statistical errors, and therefore more precise
tests of the Standard Model.

The diboson production processes presented here are
sensitive to triple gauge couplings (TGCs), which are allowed 
in the Standard Model only for specific vertices.
Anomalous triple gauge couplings, differing from those expected
in the Standard Model, could manifest as modifications of the
cross section or kinematics of diboson production.
These diboson production processes 
are also important
backgrounds in searches for the Higgs boson.
The WW and ZZ processes could happen both through decays of
a Standard Model Higgs boson, or through direct production, which
acts a large background in the search for the Higgs.

\section{The ATLAS experiment}

The ATLAS detector is a general purpose detector at the
LHC, CERN.
It consists of an inner detector for charged tracking,
surrounded by electromagnetic and hadronic calorimeters,
and finally a muon detector system.
The detector and its performance
are described in detail in~\cite{atlasdetpaper}.

\section{Event selection and cross-section measurements}

The production of a high $p_{\mathrm{T}}$ photon in association
with a W or Z boson is the highest cross-section process
considered here. As with the other diboson channels,
measurements of this process can probe anomalous TGCs.
The analysis is performed only for the case of the W/Z boson
decaying fully leptonically, and the
main backgrounds consist of jets produced in association
with a W or Z boson, where a jet fakes either a photon or
a charged lepton~\cite{wzgamma1fb}.
Events are selected by requiring one (for the W$\gamma$ case)
or two (for the Z$\gamma$ case) leptons (electron or muon, denoted $\ell$)
with 
transverse momentum ($p_{\mathrm{T}}$) of at least 15\,GeV.
In addition, an isolated photon must be reconstructed, with
transverse energy ($E_{\mathrm{T}}$) of at least 15\,GeV.
The photon must be separated from each
lepton\footnote{$\Delta R$ is the sum in quadrature of the
  separation in azimuth
  and pseudorapidity, $(\Delta R)^2 = (\Delta \phi)^2 + (\Delta \eta)^2$}
by $\Delta R>0.7$.
Finally, for the W$\gamma$ channel, missing transverse energy
($E_{\mathrm{T}}^{\textrm{miss}}$)
of at least 25\,GeV is required, and the transverse mass of
the W must exceed 40\,GeV;
for Z$\gamma$, the dilepton invariant mass must be greater
than 40\,GeV.
The results of the cross-section measurements will be
given at the end of this section.



\begin{figure}
\includegraphics[height=150pt]{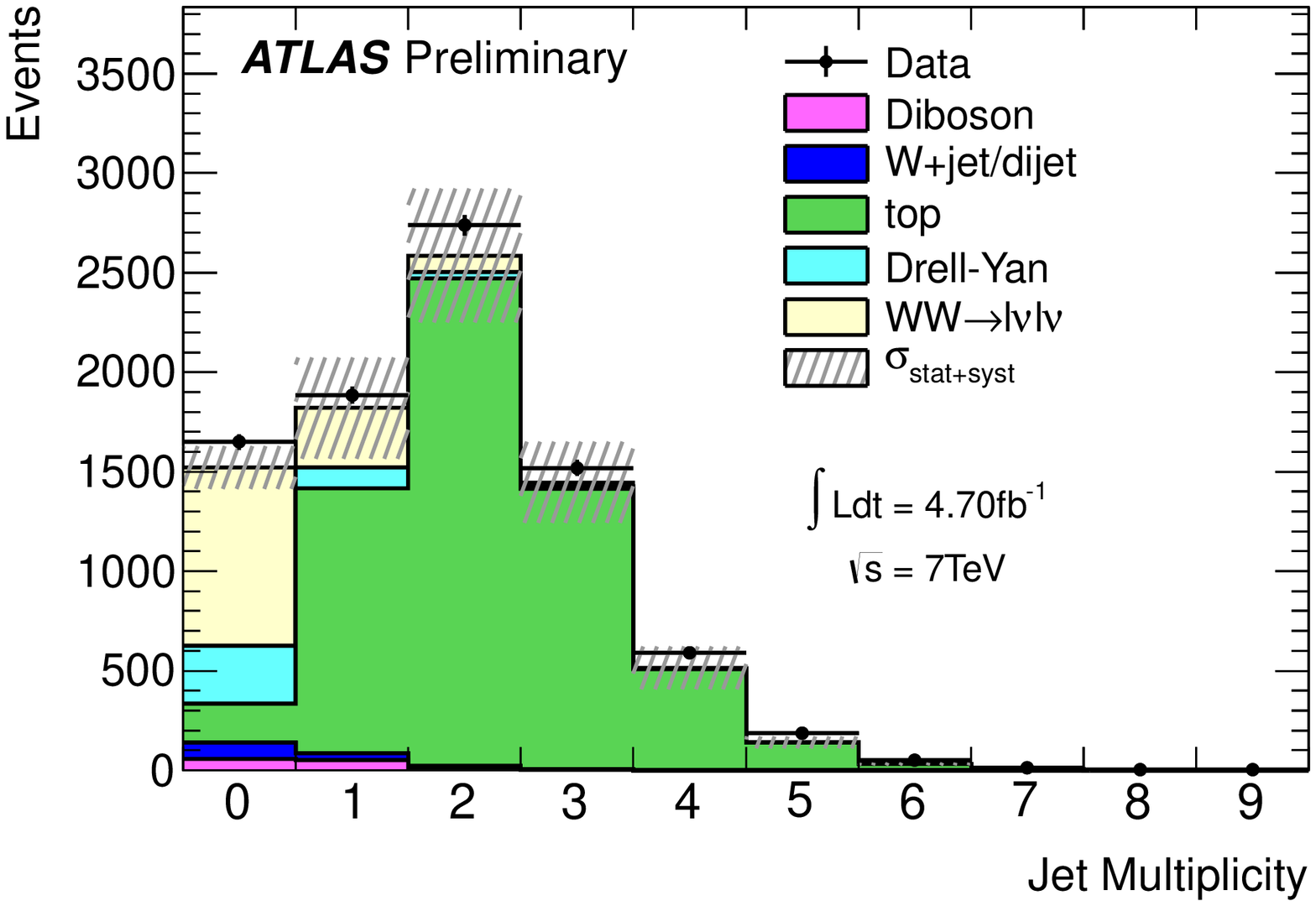}%
\hspace{\fill}%
\includegraphics[height=150pt]{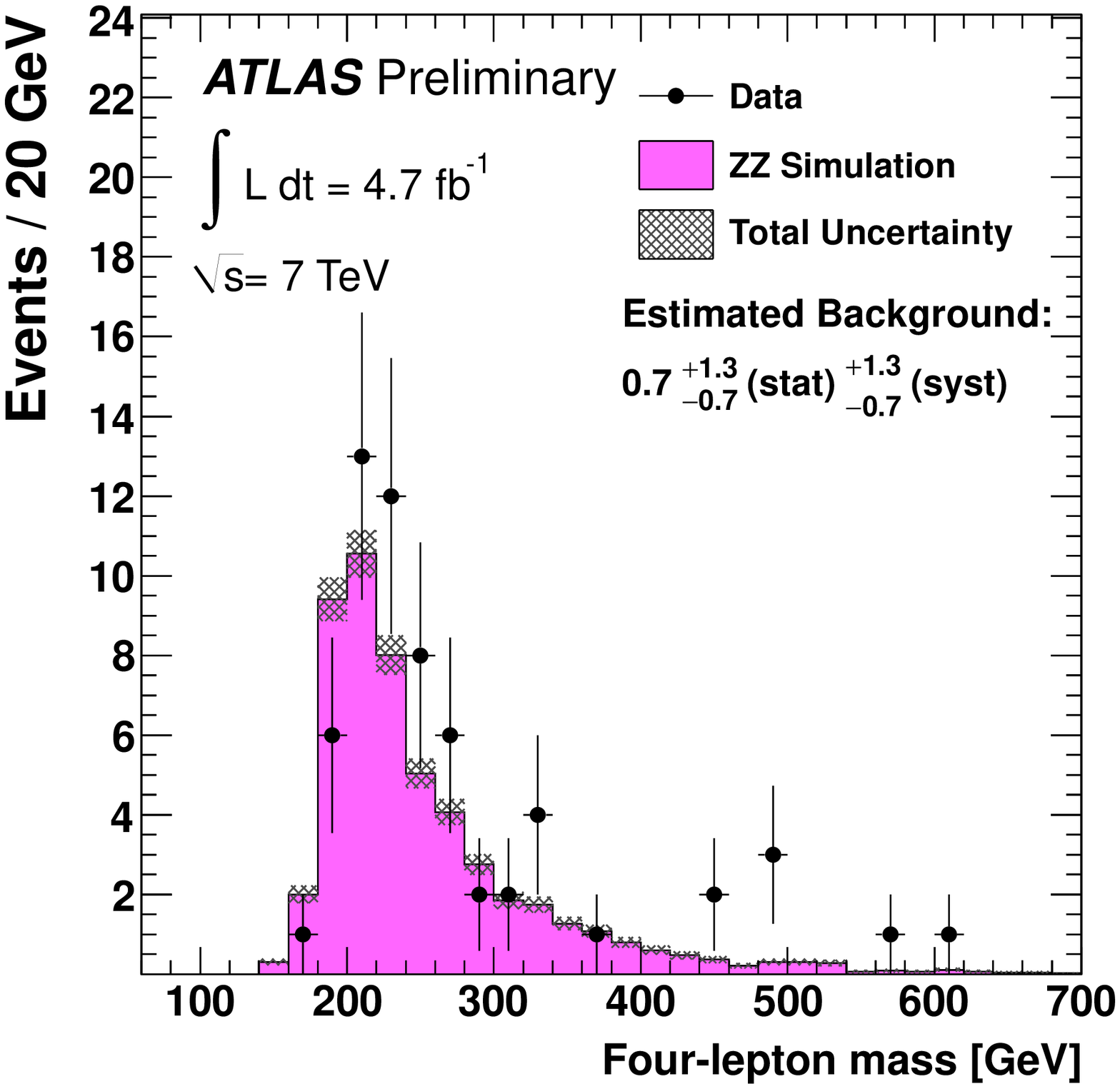}%
\\[-3pt]
\mbox{\hspace{0.28\textwidth}(a)\hspace{0.5\textwidth}(b)}

\vspace{-7pt}

\caption{(a) Jet multiplicity distribution for candidate
WW events, before the final jet veto cuts~\cite{ww5fb}
(b) Four-lepton invariant mass distribution for ZZ events
after the full selection~\cite{zzllll5fb}.
}\label{skottowe:fig:xsecone}
\end{figure}

The production of two W bosons of opposite charge, with both
W bosons decaying leptonically, is another process which has
been measured using the ATLAS dataset from 2011~\cite{ww5fb}.
This is again sensitive to anomalous TGCs, and is one of the
most important backgrounds in the search for a Standard Model
Higgs boson via the H$\to$WW decay.
The main backgrounds are the production of jets in association
with a Z boson, and top quark production.
The former is reduced by vetoing dilepton masses within 15\,GeV
of the nominal Z mass and by requiring large missing transverse
energy\footnote{We use the variable $E_{\textrm{T,rel}}^{\textrm{miss}}$,
which is the component of the $E_{\textrm{T}}^{\textrm{miss}}$ vector
which is perpendicular to the closest 
lepton or jet
if this closest object
is separated by $\Delta\phi\le\frac{\pi}{2}$;
otherwise, if $\Delta\phi>\frac{\pi}{2}$, then $E_{\textrm{T}}^{\textrm{miss}}$ is used.}; 
the latter is reduced by 
requiring events to contain
zero jets with $p_{\mathrm{T}}>25\,$GeV, and
zero jets with $p_{\mathrm{T}}>20\,$GeV that contain a b-hadron.
Figure~\ref{skottowe:fig:xsecone}(a) shows the jet multiplicity after
the 
missing transverse energy
cut, and before the jet veto cuts.



Increasing the required number of leptons from two to three,
the diboson production process WZ has also been measured~\cite{wz1fb}.
This analysis is again sensitive to anomalous TGCs, and 
searches for a charged Higgs boson.
The analysis is much cleaner than the WW analysis, with
the main backgrounds being caused by the reconstruction
of a jet as a charged lepton,
or by a true charged lepton not being detected.
These main backgrounds are production of jets in association
with a single Z boson, and ZZ diboson production,
and top quark production.
Events are selected by first requiring two isolated leptons
(either two electrons or two muons) with $p_{\mathrm{T}}>15\,$GeV,
with a dilepton mass within 10\,GeV of the nominal Z mass,
then requiring a third lepton with $p_{\mathrm{T}}>20\,$GeV 
attributed to the W.
In addition, $E_{\mathrm{T}}^{\textrm{miss}}>25$\,GeV is required,
and the transverse mass of the W boson 
is required to be greater than 40\,GeV.

Measurements have been made of ZZ diboson production in two
final states: firstly with both Z bosons decaying into charged
leptons,
and secondly with one Z decaying into charged leptons and the other
Z boson decaying to a pair of neutrinos
(sharing the final state with the WW diboson measurement described
above).
The latter final state gives a gain in branching fraction 
compared to the former, but also suffers from increased background.

For the $\ell^-\ell^+\ell^-\ell^+$ final state, four isolated 
leptons are required, with $p_{\mathrm{T}}${}$>$7\,GeV~\cite{zzllll5fb}.
The leptons are required to form two pairs, each with dilepton
invariant mass within 15\,GeV of the nominal Z mass.
The four-lepton mass distribution of selected events
is shown in figure~\ref{skottowe:fig:xsecone}(b).

For the $\ell^-\ell^+\nu^-\nu^+$ final state, two leptons with
$p_{\mathrm{T}}>$20\,GeV and with $m_{\ell\ell}$ within 15\,GeV
of the nominal Z mass are required~\cite{zzllnunu}.
Large missing transverse energy is also required.
For this the axial $E_{\mathrm{T}}^{\textrm{miss}}$ is used,
requiring the component of
the $E_{\mathrm{T}}^{\textrm{miss}}$ parallel to
the dilepton vector in the plane transverse
to the beam to be greater than 80\,GeV.
Finally, the events are required to contain zero jets with
$p_{\mathrm{T}}${}$>$25\,GeV,
and the fractional $p_{\mathrm{T}}$ difference between
$E_{\mathrm{T}}^{\textrm{miss}}$
and dilepton is required to be small:
$|E_{\mathrm{T}}^{\textrm{miss}}-p_{\mathrm{T}}^{\ell\ell}|/p_{\mathrm{T}}^{\ell\ell}${}$<$0.6.


\begin{table}
\begin{tabular}{%
@{$\mspace{7mu}$}l
@{$\mspace{10mu}$}c
@{$\mspace{20mu}$}r@{$\mspace{5mu}$}r@{$\mspace{5mu}$}r@{$\mspace{5mu}$}r
@{$\mspace{25mu}$}r@{$\mspace{5mu}$}r@{$\mspace{5mu}$}r@{$\mspace{5mu}$}r
@{$\mspace{10mu}$}c
@{$\mspace{7mu}$}%
}
Process
& $\int\mspace{-5mu}\mathcal{L}\mspace{1mu}\mathrm{d}t$
& $\sigma_{\mathrm{fid}}^{\phantom{l}}${}$\mspace{2mu}${[fb]} & {(stat.)} & {(syst.)} & {(lumi.)}
& $\sigma_{\mathrm{tot}}^{\phantom{l}}${}$\mspace{2mu}${[pb]} & {(stat.)} & {(syst.)} & {(lumi.)}
& Reference\rule[-5pt]{0pt}{10pt}\\ \hline
W$\gamma$ \rule{0pt}{9.6pt}
  & 1\,fb$^{-1}$
  & 4.60&$\pm$0.11&$\pm$0.64&$\pm$0.17
  &     &         &         &
  & \cite{wzgamma1fb}\\
Z$\gamma$ \rule{0pt}{9.6pt}
  & 1\,fb$^{-1}$
  & 1.29&$\pm$0.05&$\pm$0.15&$\pm$0.05
  &     &         &         &
  & \cite{wzgamma1fb}\\
WW \rule{0pt}{9.6pt}
  & 1\,fb$^{-1}$
  & \multicolumn{4}{c}{\footnotesize{\slshape by channel: see reference}}
  & 54.4&$\pm$4.0&$\pm$3.9&$\pm$2.0
  & \cite{ww1fb}\\
WW \rule{0pt}{9.6pt}
  & 5\,fb$^{-1}$
  & \multicolumn{4}{c}{\footnotesize{\slshape by channel: see reference}}
  & 53.4&$\pm$2.1&$\pm$4.5&$\pm$2.1
  & \cite{ww5fb}\\
WZ \rule{0pt}{11.0pt}
  & 1\,fb$^{-1}$
  &  102&$_{-14}^{+15}$&$_{-6}^{+7}$&$_{-4}^{+4}$
  & 20.5&$_{-2.8}^{+3.1}$&$_{-1.3}^{+1.4}$&$_{-0.8}^{+0.9}$
  & \cite{wz1fb}\\
ZZ$\to${}%
$\ell\mspace{1mu}\ell\mspace{1mu}\ell\mspace{1mu}\ell$
 \rule{0pt}{11.0pt}
  & 1\,fb$^{-1}$
  & 19.4&$_{-5.2}^{+6.3}$&$_{-0.7}^{+0.9}$&$\pm$0.7
  &  8.5&$_{-2.3}^{+2.7}$&$_{-0.3}^{+0.4}$&$\pm$0.3
  & \cite{zzllll1fb}\\
ZZ$\to${}%
$\ell\mspace{1mu}\ell\mspace{1mu}\ell\mspace{1mu}\ell$
 \rule{0pt}{11.0pt}
  & 5\,fb$^{-1}$
  & 21.2&$_{-2.7}^{+3.2}$&$_{-0.9}^{+1.0}$&$\pm$0.8
  &  7.2&$_{-0.9}^{+1.1}$&$_{-0.3}^{+0.4}$&$\pm$0.3
  & \cite{zzllll5fb}\\
ZZ$\to${}%
$\ell\mspace{1mu}\ell\mspace{1mu}$%
{}$\nu\nu$
 \rule{0pt}{11.0pt}
  & 5\,fb$^{-1}$
  & 12.2&$_{-2.8}^{+3.0}$&$\pm$1.9&$\pm$0.5
  &  5.4&$_{-1.2}^{+1.3}$&$_{-1.0}^{+1.4}$&$\pm$0.2
  & \cite{zzllnunu}
\end{tabular}

\vspace{-3pt}

\caption{Summary table of cross-section measurements.
Results are given both for the fiducial volume defined by the selection cuts,
$\sigma_{\mathrm{fid}}$,
and extrapolated to the total cross section, $\sigma_{\mathrm{tot}}$.}\label{skottowe:tab:xsecs}
\end{table}

The results of all the above cross-section measurements are presented in
table~\ref{skottowe:tab:xsecs}.
For each analysis, the most recent ATLAS result is given.
All measurements agree with the Standard Model prediction within the respective uncertainties.
In the case of WW and ZZ$\to${}$\ell^-\ell^+\ell^-\ell^+$,
the previous cross-section result is also given, as these have been
used to derive the TGC limits
discussed in the next section.
More details on the measurements can be found in the reference
for each analysis in the table.


\section{Limits on anomalous triple gauge couplings}

\begin{figure}
\mbox{%
{\includegraphics[viewport=2mm 0mm 184mm 152mm, clip, height=117pt]{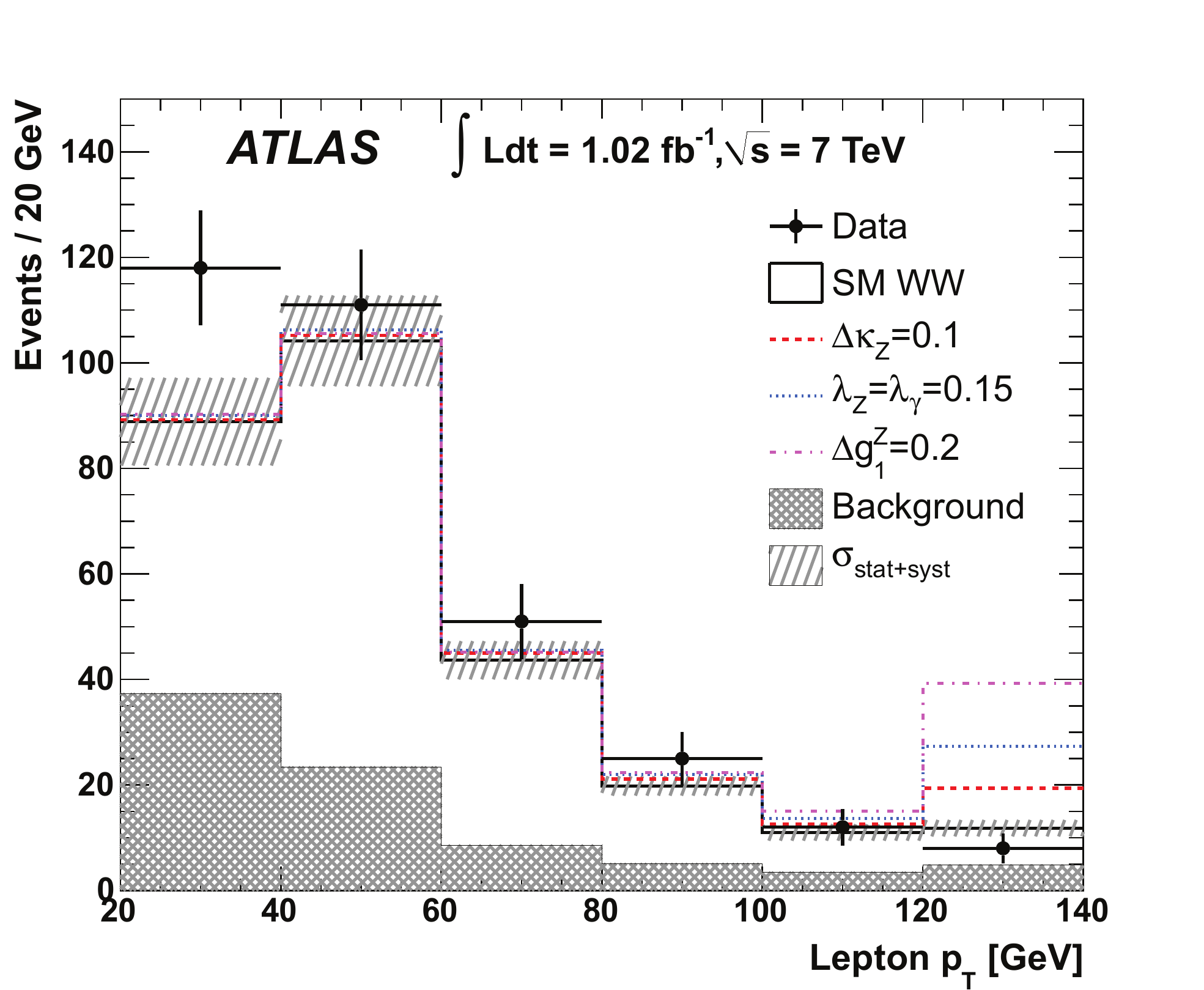}}%
\hspace{7pt}%
{\includegraphics[viewport=18mm 0mm 183mm 152mm, clip, height=117pt]{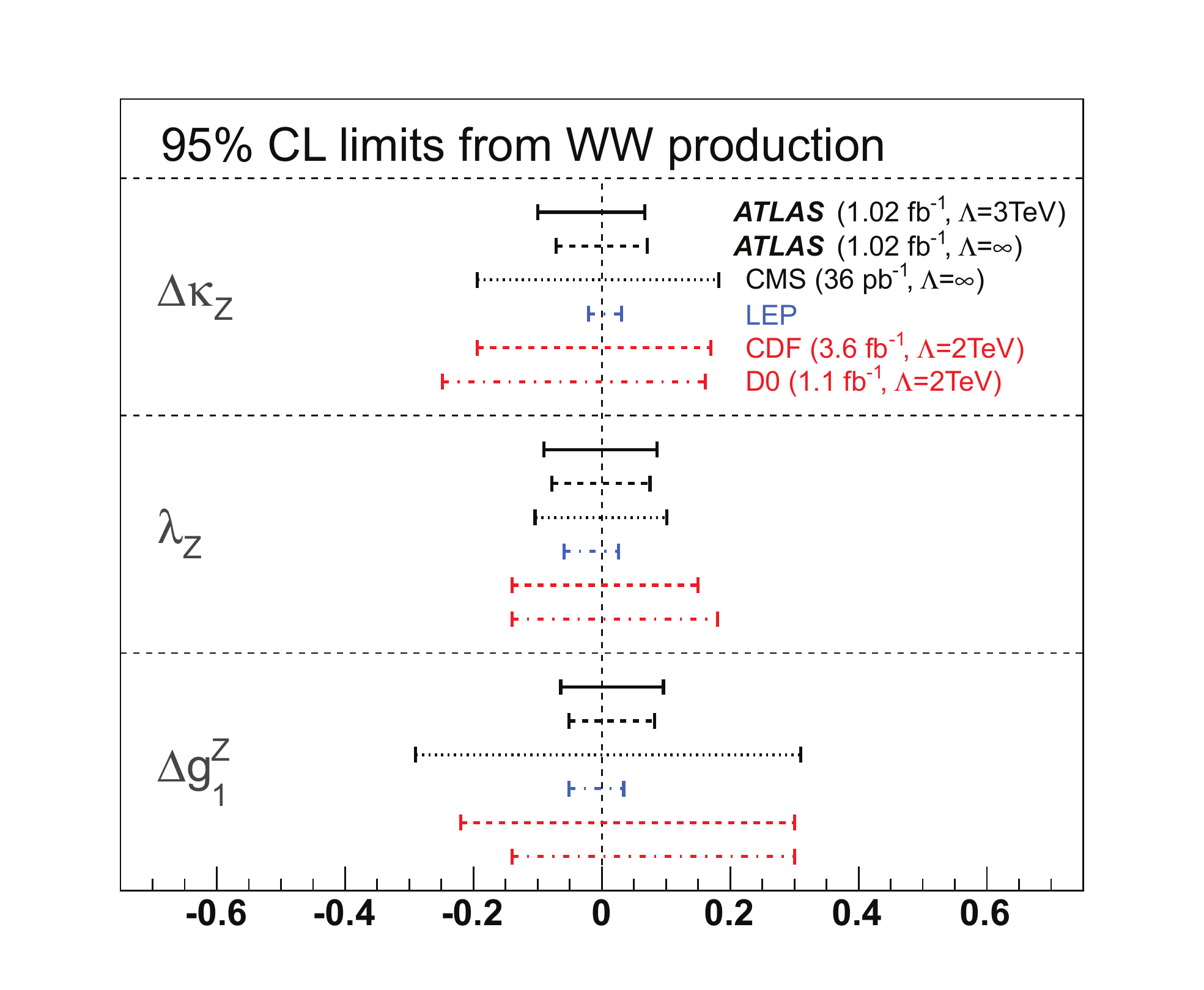}}%
\hspace{7pt}%
{\includegraphics[viewport=5mm 10mm 195mm 187mm, clip, height=117pt]{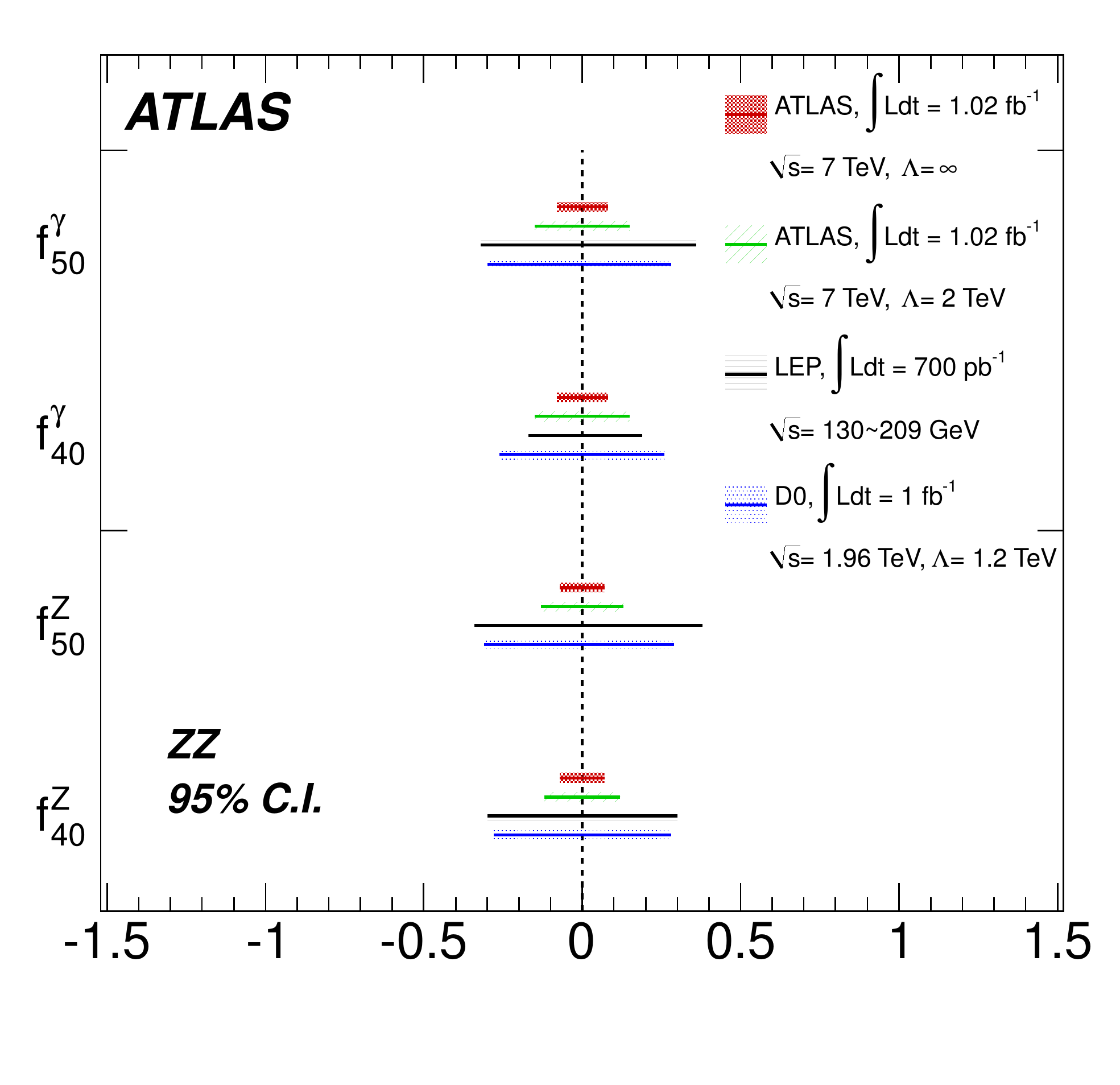}}}\\
\mbox{%
\hspace{0.17\textwidth}\rlap{(a)}%
\hspace{0.325\textwidth}\rlap{(b)}%
\hspace{0.33\textwidth}\rlap{(c)}}


\caption{(a) Leading-lepton $p_{\mathrm{T}}$
spectrum of WW events, 
for data and
MC with various TGCs
($\Delta\kappa_{\mathrm{z}}$,
$\lambda_{\mathrm{z}}$ and
$\Delta g_{1}^{\mathrm{z}}$
are all equal to zero in the SM)~\cite{ww1fb}.
This spectrum is used to derive the limits on anomalous TGCs shown in (b).
(c) Limits on anomalous TGCs computed from the cross-section measurement of
ZZ$\to${}$\ell^-\ell^+\ell^-\ell^+$~\cite{zzllll1fb}.
}\label{skottowe:fig:atgcone}
\end{figure}

The final selected events in each of the analyses described above can
be used further, to derive limits on anomalous triple gauge couplings (TGCs).
Anomalous TGCs can result in deviations of the cross sections
and kinematics of these processes, and therefore cross-section measurements,
or kinematic distributions, can be used to impose limits on these couplings.
Figure~\ref{skottowe:fig:atgcone}(a) shows the example of the WW analysis,
where the leading-lepton $p_{\mathrm{T}}$ spectrum is very sensitive to anomalous
TGCs, particularly at large values. The distribution observed in data is
used to compute the limits on anomalous TGCs shown in figure~\ref{skottowe:fig:atgcone}(b).
Figure~\ref{skottowe:fig:atgcone}(c) shows the limits computed using 
ZZ$\to${}$\ell^-\ell^+\ell^-\ell^+$ events.
Anomalous TGC limits have also been derived using
W$\gamma$, Z$\gamma$~\cite{wzgamma1fb} and
WZ~\cite{wz1fb} events.



\section{Summary}

Cross sections have been measured for a number of diboson production
processes, both extrapolated to the total phase space, and within
a fiducial volume given by the detector acceptance and event selection cuts.
These measurements have been made for production of W$\gamma$, Z$\gamma$,
WW, WZ and ZZ,
using part or all of the ATLAS dataset from the
2011 run of the LHC at $\sqrt{s}=7\,$TeV.
These analyses form precise tests of the Standard Model, and have been
used to place stringent limits on anomalous triple
gauge couplings.










{\raggedright
\begin{footnotesize}



\end{footnotesize}
}


\end{document}